\def\aj{{\it AJ}}  
\def\apj{{\it ApJ}}  
\def\pasp{{\it PASP}}      
\def\mnras{{\it MNRAS}}      

\def\aap{{\it A\&A}}     
\def\aapr{{\it {A\&A~Rev.}} }
\def\araa{{\it {ARA\&A}} }
\documentclass{iaus}
\usepackage{graphicx}

\title[Activity Ages of M Dwarfs] 
{Using Magnetic Activity and Galactic Dynamics to Constrain the Ages of M Dwarfs}

\author[West et al.]   
{Andrew A. West$^1$
, Suzanne L. Hawley$^2$, John J. Bochanski$^1$, Kevin R. Covey$^3$ \and Adam J. Burgasser$^1$}

\affiliation{$^1$MIT Kavli Institute for Astrophysics and Space Research,\\ 77
  Massachusetts Ave, Cambridge, MA 02139-4307 \\ email: {\tt aaw@mit.edu, jjb@mit.edu, ajb@mit.edu} \\[\affilskip]
$^2$Department of Astronomy, University of Washington,\\ Box 351580,
Seattle, WA 98195 \\email: {\tt slh@astro.washington.edu}
\\[\affilskip]
$^3$Harvard-Smithsonian Center for Astrophysics,\\ 60
  Garden Street, Cambridge MA 02138  \\ email: {\tt kcovey@cfa.harvard.edu}}

\pubyear{2008}
\volume{258}  
\pagerange{}
\jname{The Ages of Stars}
\editors{E. Mamajek, D. Soderblom \& J. Valenti, eds.}
\begin{document}

\maketitle

\begin{abstract}
  We present a study of the dynamics and magnetic activity of M dwarfs
  using the largest spectroscopic sample of low-mass stars ever
  assembled. The age at which strong surface magnetic activity (as
  traced by H$\alpha$) ceases in M dwarfs has been inferred to have a
  strong dependence on mass (spectral type, surface temperature) and
  explains previous results showing a large increase in the fraction
  of active stars at later spectral types. Using spectral observations
  of more than 40000 M dwarfs from the Sloan Digital Sky Survey, we
  show that the fraction of active stars decreases as a function of
  vertical distance from the Galactic plane (a statistical proxy for
  age), and that the magnitude of this decrease changes significantly
  for different M spectral types. Adopting a simple dynamical model
  for thin disk vertical heating, we assign an age for the activity
  decline at each spectral type, and thus determine the activity
  lifetimes for M dwarfs. In addition, we derive a statistical
  age-activity relation for each spectral type using the dynamical
  model, the vertical distance from the Plane and the H$\alpha$
  emission line luminosity of each star (the latter of which also
  decreases with vertical height above the Galactic plane).
  \keywords{stars: activity, stars: kinematics, stars:
    late-type,stars: low-mass, brown dwarfs}
\end{abstract}

\firstsection
\section{Introduction}

M dwarfs are the smallest and least luminous stars on the main
sequence, yet they are the most numerous of all stellar constituents
and have lifetimes longer than the current age of the
Universe. Determining the age of an individual main sequence, field M
dwarf is quite challenging.  Stellar clusters would seem to provide
the ideal environments for calibrating the ages of M dwarfs (because
the age of the cluster is determined). However, the large distances to
clusters older than a few Gyr (and resulting faintness of cluster M
dwarfs), preclude any detailed observations of the intermediate or
oldest cluster resident M dwarfs. Age estimates for field M dwarfs
must therefore rely on a number of techniques that piggyback on other
age dating methods.  Although several methods exist for estimating the
ages of M dwarfs (e.g. contribution by S. Catalan), we focus this
contribution on what the magnetic activity of M dwarfs can tell us
about their age.

M dwarfs are host to intense magnetics dynamos that give rise to
chromospheric and coronal heating, producing emission from the X-ray
to the radio. This magnetic activity has long been thought to be tied
to the age of the host star.  Almost 40 years ago, Wilson \& Woolley
(1970) found a link between magnetic activity (as traced by the Ca II
H \& K emission lines) and the orbital elements (namely the
inclination and eccentricity) of more than 300 late-type dwarfs; stars
in near-circular orbits with small inclinations had the strongest
activity.  They concluded that as stars age, their orbits get more
inclined and more eccentric (due to dynamical interactions), and thus,
magnetic activity must also decline with age.  Subsequent studies over
the following decades have found similar connections between age and
activity in low-mass stars (Wielen 1977; Giampapa \& Liebert 1986;
Soderblom et al. 1991; Hawley et al. 1996, Hawley et al. 1999, Hawley
et al. 2000).

Although activity has been observed for decades, the exact mechanism
that gives rise to the chromospheric heating is still not
well-understood.  In the Sun, magnetic field generation and subsequent
heating is closely tied to the Sun's rotation.  From helioseismology,
we know that there is a rotational boundary between the inner
solid-body rotating radiative zone and the outer differentially
rotating convective zone (Parker 1993; Ossendrijver 2003; Thompson et
al. 2003).  This boundary, dubbed the tachocline, creates a rotational
shear that likely allows magnetic fields to be generated, preserved
and eventually rise to the surface where they emerge as magnetic
loops.  These loops bring heat to the Sun's chromosphere and corona,
driving both large stellar flares, as well as lower-level quiescent (or
persistent) magnetic activity.  The faster a star (with a tachocline)
rotates, the stronger its magnetic heating and surface activity.
Therefore, the angular momentum evolution of solar-type stars should
play an important role in determining the magnitude of the observed
activity.

Angular momentum loss from magnetized winds has been shown to slow
rotation in solar-type stars; as a result magnetic activity decreases.
Skumanich (1972) found that both activity and rotation decrease over
time as a power law ($t^{-0.5}$).  Subsequent studies confirmed the
Skumanich results and empirically demonstrated a strong link between
age, rotation and activity in solar-type stars (Barry 1988; Soderblom
et al. 1991; Pizzolato et al. 2003; Mamajek \& Hillenbrand 2008; see
contribution by E. Mamajek).

There is strong evidence that the rotation-activity (and presumably
age) relation extends from stars more massive than the Sun to smaller
dwarfs (Pizzolato et al. 2003; Mohanty \& Basri 2003; Kiraga \& Stepien
2007). Therefore, rotation derived ages (using gyrochronology) may
provide a useful independent estimate of age for M dwarfs when large
enough calibration samples can be acquired (see contributions by
S. Meibom, J. Irwin, and S. Barnes).  However, at a spectral type of
$\sim$M3 (0.35 M$_{\odot}$; Reid \& Hawley 2005; Chabrier \& Baraffe
1997), stars become fully convective and the tachocline presumably
disappears.  Despite this changes, magnetic
activity persists in late-type M dwarfs; the fraction of active M
dwarfs peaks around a spectral type of M7 before decreasing into the
brown dwarf regime (Hawley et al. 1996; Gizis et al. 2000; West et
al. 2004).

It is unclear if the rotation-age-activity relation extends to the
fully convective M dwarfs.  A few empirical studies have uncovered
evidence that activity and rotation might be linked in late-type M
dwarfs (Delfosse et al. 1998; Mohanty \& Basri 2003; Reiners \& Basri
2007).  In addition, recent simulations of fully convective stars find
that rotation may play a role in magnetic dynamo generation (Dobler et
al. 2006; Browning 2008).  However, a recent study found detectable
rotation in a few inactive M7 dwarfs, indicating that the situation
may in fact be more complicated (West \& Basri 2009).

Irrespective of rotation, many studies have found evidence that the
age-activity relation extends into the M dwarf regime.  Eggen (199)
observed a Skumanich-type decrease in activity as a function of age.
Larger samples of M dwarfs have added further evidence that magnetic
activity decreases with age (Fleming et al. 1995; Gizis et al. 2002).
There is also evidence that M dwarfs may have finite active lifetimes.
Stauffer et al. (1994) suggested that activity may not be present in
the most massive M dwarfs in the Pleiades.  Hawley et al. (2000)
confirmed the Stauffer et al. (1994) claim by observing a a sample of
clusters that spanned several Gyr in age.  They were able to calibrate
the lifetimes for early-type M dwarfs by observing the color at which
activity (traced by H$\alpha$ emission) was no longer present.
Because of the small sample size, the derived Hawley et al. (2000)
activity ages are lower limits at a given color or spectral type. As
mentioned above, the Hawley et al. (2000) study was limited to
younger, nearby clusters due to the intrinsic faintness of M dwarfs;
thus, it could only probe ages of a few Gyr and spectral types as late
as $\sim$M3. Larger samples of M dwarfs are required to statistically
derive age-activity relations that extend both in the ages they probe
and the range of spectral types they cover.

In this contribution, we review recent studies that have used large
spectroscopic samples to investigate the relationship between age and
activity in M dwarfs.  In addition, we derive a statistical H$\alpha$
age-activity relation for M2-M7 dwarfs.

\section{SDSS DR5 Low-Mass Star Spectroscopic Sample}

Large surveys such as the Sloan Digital Sky Survey (SDSS; York et
al. 2000) have created optical and infrared catalogs of tens of
millions of M dwarfs (Bochanski et al. 2009 in prep).  In addition to
the photometric data, the SDSS has obtained spectra for tens of
thousands of M dwarfs.  Recently, these surveys have been utilized to
examine the statistical properties of M dwarfs, the dynamics of the
Milky Way and detailed studies of magnetic activity (Hawley et
al. 2002; West et al. 2004; West et al. 2006; Covey et al. 2007;
Bochanski et al. 2007a, 2007b; West et al. 2008; hereafter W08).

Recently, the SDSS Data Release 5 (DR5) low-mass star spectroscopic
sample was released (W08)\footnote{Measured quantities can be obtained
  electronically using the CDS Vizier database
  http://vizier.u.strasbg.fr/vis-bin/VizieR}, containing over 44000 M
and early L-type dwarfs.  Radial velocities (accurate to within
$\sim$5 km s$^{-1}$) and photometric distances (from the $r-z$ colors;
see Bochanski et al. 2009 in prep) were measured for all of the
stars. Proper motions (matches to USNO-B; Munn et al. 2004) were
obtained for over 27000 of the stars, allowing for full $U$, $V$, $W$
3-D space motions for over 27000 low-mass dwarfs, the largest such
sample ever assembled.  In addition, all of the spectra were run
through the Hammer spectral typing facility (Covey et al. 2007) to
assign spectral types and measure the strength of the H$\alpha$
emission line.  One of the hallmarks of the W08 sample is that it
probes the entirety of the Galactic thin disk and extends well into
the thick disk ($\sim$50-2500 pc above the Plane).  This is due to the
fact that the majority of SDSS fields are in the North Galactic Cap
and that SDSS was designed to study the distant Universe;
intrinsically faint Galactic M dwarfs are bright in a deep
extragalactic survey.

\section{Results}

\subsection{H$\alpha$ Activity}
\begin{figure}[h]
\begin{center}
 \includegraphics[width=3.0in]{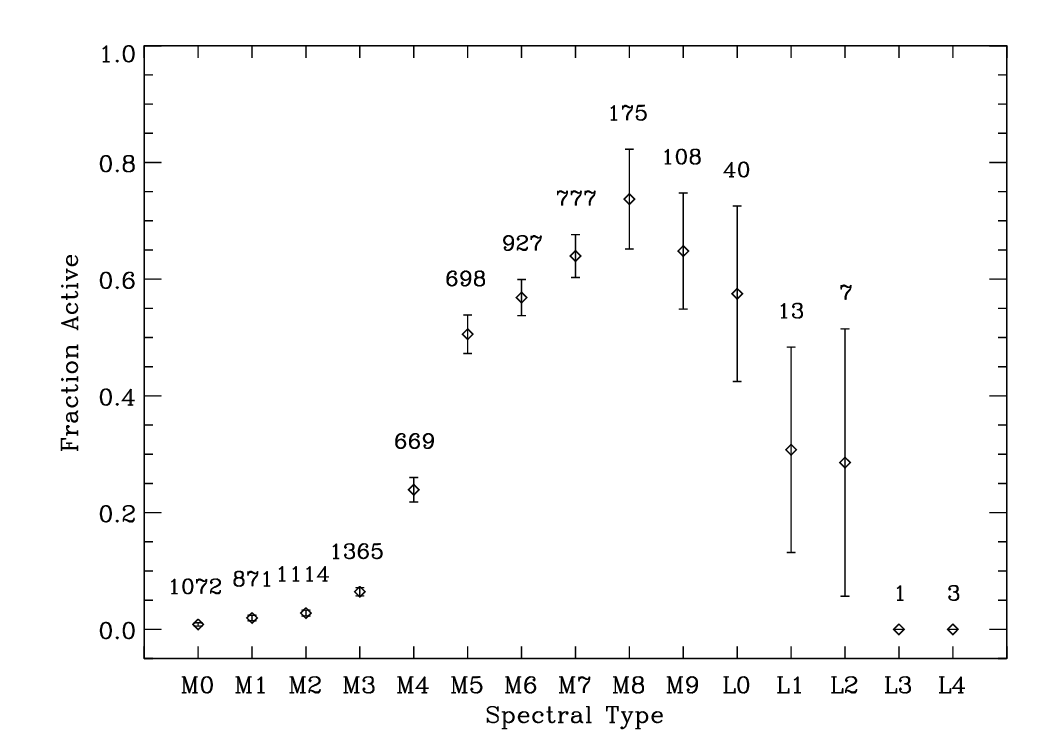} 
 \caption{Fraction of active stars as a function of spectral type
   (reproduced from West et al. 2004).  Numbers above each point
   represent the number of stars used to compute the fraction. }
   \label{frac}
\end{center}
\end{figure}

Activity in M dwarfs varies as a function of spectral type.  Figure
\ref{frac} shows the fraction of active M dwarfs as a function of
spectral type from West et al. (2004). The active fraction is very
small for early-type M dwarfs, peaks at M7-M8 and declines into the L
dwarf regime.  While some of the morphology of this relation may be
due to the ability (or inability) to host a strong dynamo (the
activity fraction decrease in the late-type M and L dwarfs is likely
caused by the atmospheres becoming neutral), a large part of the shape
may be due to age effects.  If activity has a finite lifetime that
changes as a function of spectral type, and we observe a range of
stellar ages, then stars with longer lifetimes will appear to have
higher activity fractions.

\begin{figure}[h]
\begin{center}
 \includegraphics[width=3.0in]{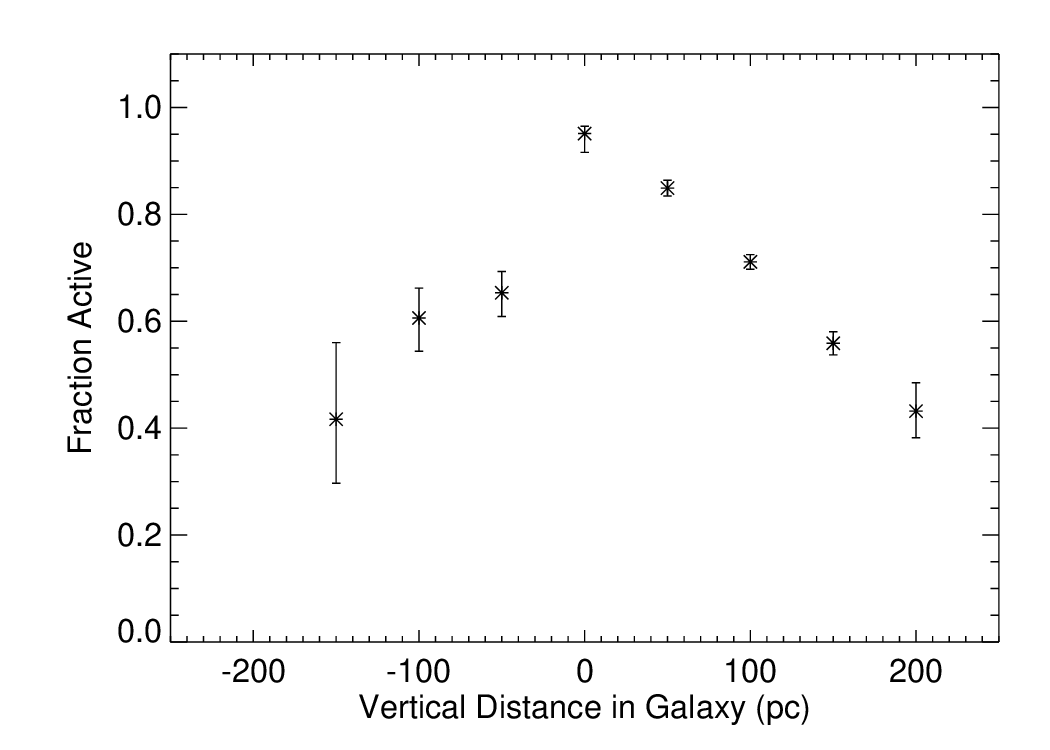} 
 \caption{The fraction of active M7 stars as a function of vertical
   distance from the Galactic plane (reproduced from West et
   al. 2006). There is a significant decrease in the active fraction as
   a function of Galactic height, which can be used as a proxy for
   age. Younger stars near the Plane are almost all active,
   whereas older stars, further from the Plane have ceased being
   active.}
   \label{falloff}
\end{center}
\end{figure}

To test this hypothesis, West et al. (2004, 2006) showed that the
fraction of active M7 stars decreases as a function of vertical
distance from the Galactic Plane (see Figure \ref{falloff}).  Stars
are born in dynamically cold molecular gas near the midplane of
the Galaxy.  Over time they undergo dynamical interactions with
molecular clouds, which add energy to the stellar orbits in all
dimensions (see contribution by B. Nordstr{\"o}m).  It is this process
that gives thickness to the Milky Way disk and allows us to use
vertical distance as a proxy for age; stars further from the Plane are
statistically older (they have to be dynamically heated to reach those
heights), while stars near the Plane are statistically younger.  Figure
\ref{falloff} shows that the younger stars, near the plane of the Milky
Way are almost all active, whereas the fraction of active stars at
larger distances above (and below) the Galaxy is significantly
reduced; older stars have ceased being active.  The decrease of active
fractions as a function of vertical height above the Plane is seen
for all M dwarf spectral types (W08).

\begin{figure}[h]
\begin{center}
 \includegraphics[width=3.5in]{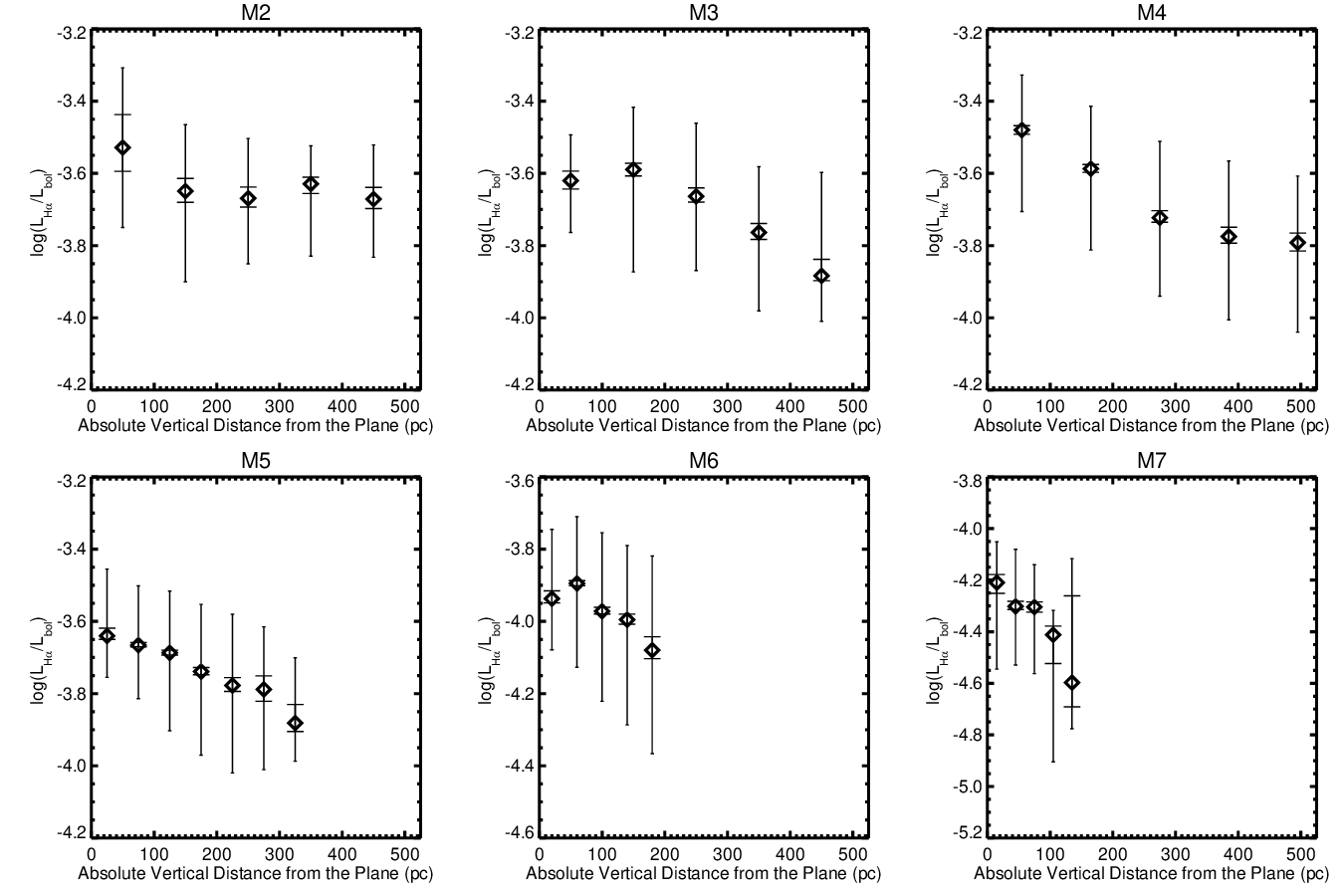} 
 \caption{Median log($L_{\rm{H}\alpha}$/$L_{\rm{bol}}$) as a function of
   vertical distance from the Plane (reproduced from W08).  The narrow
   error bars represent the spread of the values and the wide error
   bars indicate the uncertainty of the median relation in each bin.
   The decrease as a function of height is a statistically significant
   over most of the spectral types.}
   \label{lbol}
\end{center}
\end{figure}

W08 also showed that the amount of activity, quantified by
$L_{\rm{H}\alpha}$/$L_{\rm{bol}}$ (the ratio of the luminosity in the
H$\alpha$ emission line to the bolometric flux of the star) decreases
over time.  Figure \ref{lbol} shows the median
log($L_{\rm{H}\alpha}$/$L_{\rm{bol}}$) as a function of vertical
distance from the Plane.  The narrow error bars represent the spread
of the values and the wide error bars indicate the uncertainty of the
median relation in each bin.  The decrease as a function of height is
a statistically significant over most of the spectral types.  Some of
the spectral types were omitted from Figure \ref{lbol} because they
lacked a sufficient number of active stars for a robust study.

\subsection{1-D Dynamical Model}

West et al. (2006, 2008) devolved a 1D dynamical model that traces the
vertical dynamics of stars as a function of time.  The model assumes a
constant star formation rate and adds a new population of 50 stars
every 200 Myr for a total simulation time of 10 Gyr.  Each new
population of stars begins with a randomly drawn velocity dispersion 8
km s$^{-1}$ (Binney et al. 2000).  Orbits are integrated using a
``leap frog'' integration technique (Press et al. 1992) and the
vertical Galactic potential from Siebert et al. (2003).

Dynamical heating was simulated by altering the velocities of stars
such that their velocity dispersions as a function of age match a
$\sigma_W \propto t^{0.5}$ relation (Wielen 1977; Fuchs et al. 2001;
H{\"a}nninen \& Flynn 2002; Nordstr{\"o}m et al. 2004).  Energy was added to
stars that were within a certain distance from the Plane.  This
``region of influence'' is a way to parameterize the cross section of
interaction with molecular clouds during a Plane crossing (see West et
al. 2006 for more information). The majority of the molecular gas is
constrained to a small range of Galactic heights and
the dynamical interaction depends on the proximity to the cloud, the
density of the gas and the velocity of the star, all of which are
absorbed in the ``region of influence''.  The size of the ``region of
influence'' was varied from $\pm$0.5 to $\pm$5 pc in intervals of 0.5
pc.  Seventy simulations were run for each spectral type.  Each
simulation recorded the velocity, position, and age of each star.

\subsection{Age-Activity}

\begin{figure}[h]
\begin{center}
 \includegraphics[width=3.1in]{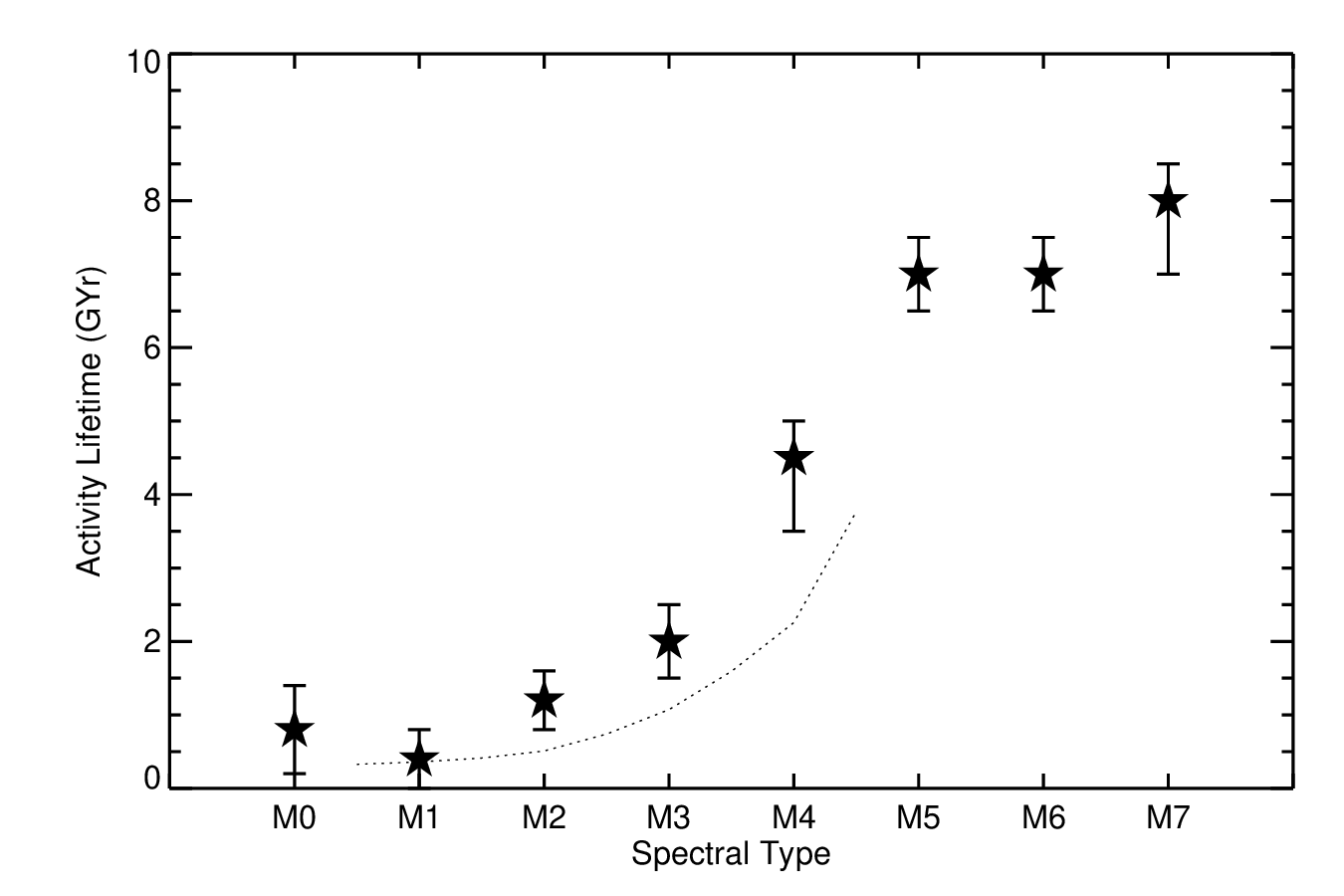} 
 \caption{The H$\alpha$ activity lifetimes of M dwarfs as determined
   by comparing the SDSS DR5 spectroscopic data to 1D dynamical
   simulations (stars; reproduced from W08).  The Hawley et al. (2000)
   activity lifetime relation is overplotted for comparison.  As
   predicted, the Hawley relation provides a lower limit to the ages.
   W08 found that there is a significant increase in activity
   lifetimes between spectral types M3 and M5, possibly indicating a
   physical change in the production of magnetic fields as full
   convection sets in. }
   \label{lifetimes}
\end{center}
\end{figure}

W08 introduced a binary activity state to the dynamical models; stars
started their lives as active and after a finite ``lifetime'', the
activity turned off.  The active lifetimes were varied from 0.0 to 9.0
Gyr in 0.5 Gyr intervals for each spectral type.  The resulting model
active fractions (as a function of vertical distance from the Plane)
were compared to the empirical relations using a chi-squared
minimization technique and a best-fit model for each spectral type was
determined.  The resulting H$\alpha$ activity lifetimes are shown in
Figure \ref{lifetimes}.  The results from the Hawley et al. (2000)
cluster study are overplotted for comparison (dotted). W08 found that
there is a significant increase in activity lifetimes between spectral
types M3 and M5, possibly indicating a physical change in the
production of magnetic fields as the stellar interiors become fully
convective.

\begin{figure}[h]
\begin{center}
 \includegraphics[width=3.6in]{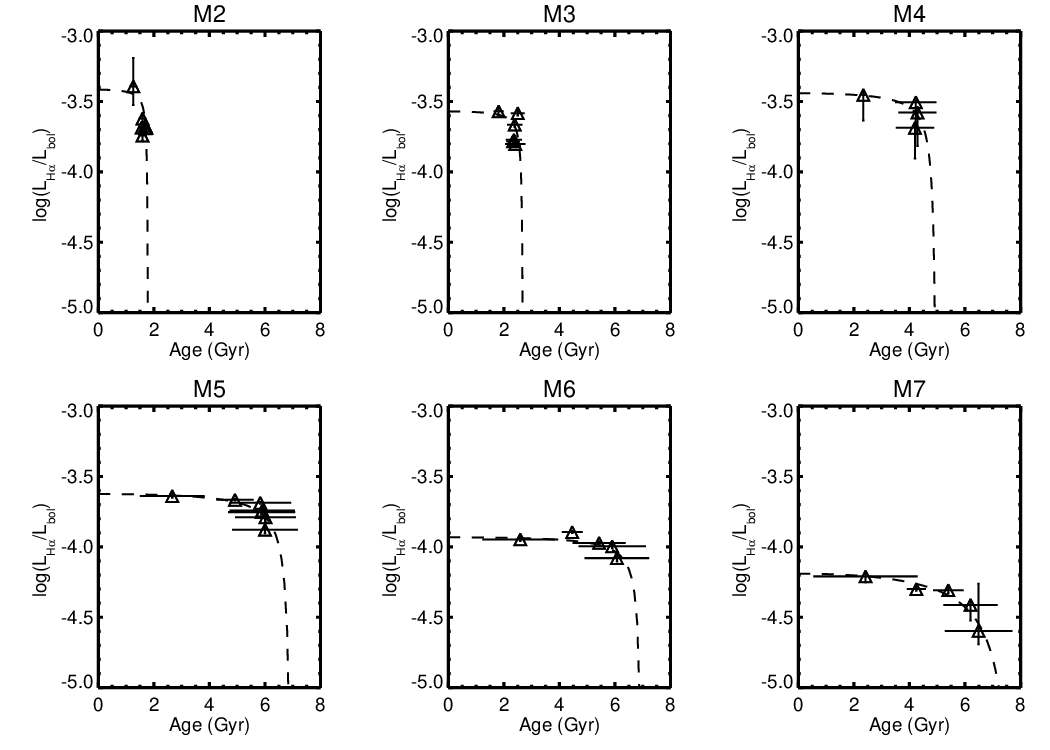} 
 \caption{Median log($L_{\rm{H}\alpha}$/$L_{\rm{bol}}$) as a function
   of age (as derived from the median age of the same bins in the W08
   dynamical models). The error bars represent the spread in the
   distributions.  The best-fit function is overplotted for each
   spectral type (dashed).}
   \label{actage}
\end{center}
\end{figure}

Figure \ref{lbol} suggests that activity is not simply a binary
process, but rather that the amount of magnetic activity may decrease
over time.  To quantify this decrease, we compared the median
log($L_{\rm{H}\alpha}$/$L_{\rm{bol}}$) from Figure \ref{lbol} to the
median ages in the same vertical distance bins from the W08 1D
dynamical models. The result, shown in Figure \ref{actage}, gives a
statistical, quantitative relationship between the H$\alpha$ activity
of an M dwarf and it's age.  Horizontal error bars represent the
spread in age at a given bin.  The age-activity relations are
consistent with a smooth Skumanich-like activity decrease until the
star reaches its activity lifetime, after which it rapidly falls to
its eternal inactive state.

We fit the activity-age relations in Figure \ref{actage} with a function of the form:

\begin{equation}
{\rm{log}}(L_{\rm{H\alpha}}/L_{\rm{bol}})=\frac{a}{l^n-t^n} - b,
\end{equation}

\noindent where $a$, $b$ and $n$ are coefficients, $l$ is the active lifetime (Gyr), and $t$ is
the age measured in Gyr. The exponent $n$ was forced to be the same for
all spectral types. The lifetime $l$ coefficient was allowed to vary
within the uncertainties of the derived active lifetimes from W08
(Figure \ref{lifetimes}). The resulting best-fit coefficients can be found
in Table \ref{coef}.  We caution that these relations are purely
statistical and may not be appropriate for use with individual stars.
Additional discretion should be used with the functional fits, which
may not be justifiably extrapolated beyond where data exist.

\begin{table}[h]
  \begin{center}
\caption{Coefficients for best-fit activity-age relation}
\label{coef}
{\scriptsize
\begin{tabular}{ccccc}\hline
{\bf Spectral Type} & {\bf $a$} &{\bf  $b$} & {\bf $n$} & {\bf $l$ (Gyr)} \\ \hline
M2 & 0.106 & 3.38 & 2.0 & 1.8\\
M3 & 0.213 & 3.54 & 2.0 & 2.7\\
M4 & 1.41 & 3.39 & 2.0 & 5.0\\
M5 & 2.85 & 3.57 & 2.0 & 7.0\\
M6 & 1.78 & 3.90 & 2.0 & 7.0\\
M7 & 11.8 & 4.01 & 2.0 & 8.0\\ \hline
\end{tabular}
}
\end{center}
\end{table}

\section{Summary}

Large astronomical surveys have produced M dwarf samples of
unprecedented size.  We have shown that using the statistical foothold
of the largest low-mass spectroscopic sample ever assembled, a strong
tie between magnetic activity (as traced by H$\alpha$) and age has
been established and that M dwarfs have finite active lifetimes.
These lifetimes are a strong function of spectral type, dramatically
increasing as the M dwarf interiors become fully convective.  We
derived age-activity relations for M2-M7 dwarfs based on the average
activity as a function of height above (or below) the Galactic plane.
Future studies will extend these relations to other spectral types.
In addition, these relations can be confirmed and calibrated using
both white dwarf cooling ages in wide binary systems that have both a white
dwarf and an M dwarf (see contributions by S. Catalan, M. Salaris, and
J. Kalirai; Silvestri et al. 2006), and deep cluster observations of M
dwarfs (when such observations become possible).


\nocite{*}

\end{document}